\documentclass[conference]{IEEEtran}
\IEEEoverridecommandlockouts
\usepackage{cite}
\usepackage{amsmath,amssymb,amsfonts}
\usepackage{algorithmic}
\usepackage{graphicx}
\usepackage{textcomp}
\usepackage{xcolor}
\usepackage{tcolorbox}
\usepackage{comment}
\usepackage{soul,color}
\usepackage{xcolor}
\usepackage{hyperref}
\usepackage{todonotes}
\usepackage{dblfloatfix} 
\usepackage{soul}
\usepackage{varwidth}
\usepackage{tikz}
\usepackage{balance}

\setuptodonotes{fancyline, color=blue!30, size=\footnotesize}

\tcbuselibrary{breakable}
\tcbuselibrary{skins}
\NewTotalTColorBox[auto counter]{\Lesson}{ +m }{ 
    breakable,
    notitle,
    colback=green!3!white,
    frame hidden,
    boxrule=0pt,
    enhanced,
    sharp corners,
    borderline west={4pt}{0pt}{green!20!white},
    boxsep=2pt,  % Decrease padding around the content
    left=7pt,    % Reduce left inner padding
    right=2pt,   % Reduce right inner padding
    top=2pt,     % Reduce top inner padding
    bottom=2pt,  % Reduce bottom inner padding
}{
    \textcolor{green!70!black}{
        \sffamily
        \textbf{Lesson~\thetcbcounter.}
    }%
    #1
}

\def\BibTeX{{\rm B\kern-.05em{\sc i\kern-.025em b}\kern-.08em
    T\kern-.1667em\lower.7ex\hbox{E}\kern-.125emX}}

\newcommand{\threatlabel}[1]{%
  \tikz[baseline=(t.base)]{
    \node[draw=red!50!black,
          rounded corners=2pt,
          fill=red!10!white,       
          inner sep=2pt,
          font=\sffamily\footnotesize
          ] (t) {#1};
  }%
}

\newcommand{\mitigation}[1]{%
  \tikz[baseline=(t.base)]{
    \node[draw=green!50!black,
          rounded corners=2pt,
          fill=green!10!white,       
          inner sep=2pt,
          font=\sffamily\footnotesize
          ] (t) {#1};
  }%
}

\begin{document}

\title{Security-by-Design at the Telco Edge with OSS: \\ Challenges and Lessons Learned}

\author{
    \IEEEauthorblockN{Carmine Cesarano, Alessio Foggia, Gianluca Roscigno, Luca Andreani, Roberto Natella}
    
    \IEEEauthorblockA{\textit{Università degli Studi di Napoli Federico II, Naples, Italy}}

    \{carmine.cesarano2, alessio.foggia, roberto.natella\}@unina.it
    \hl{aggiungere autori dalle aziende}
}

\author{
    \IEEEauthorblockN{Carmine Cesarano\textsuperscript{1}, Alessio Foggia\textsuperscript{1}, Gianluca Roscigno\textsuperscript{2}, Luca Andreani\textsuperscript{3}, Roberto Natella\textsuperscript{1}}
    
    \IEEEauthorblockA{\textit{\textsuperscript{1}Università degli Studi di Napoli Federico II, Naples, Italy}}
    
    \IEEEauthorblockA{\textit{\textsuperscript{2}System Management S.p.A, Naples, Italy}}
    \IEEEauthorblockA{\textit{\textsuperscript{3}DigitalPlatforms S.p.A, Rome, Italy}}

}

\maketitle

\begin{abstract}
This paper presents our experience, in the context of an industrial R\&D project, on securing GENIO, a platform for edge computing on Passive Optical Network (PON) infrastructures, and based on Open-Source Software (OSS). We identify threats and related mitigations through hardening, vulnerability management, digital signatures, and static and dynamic analysis. In particular, we report lessons learned in applying these mitigations using OSS, and share our findings about the maturity and limitations of these security solutions in an industrial context.
\end{abstract}


\section{Introduction}

% Edge-PON integration 
Edge computing has emerged as a transformative paradigm in the telecommunications and industrial sectors, enabling low-latency data processing closer to the end-users \cite{Ahmed2017BringingCC}. 
Traditionally, edge computing and broadband access networks have operated independently, with edge workloads running on dedicated servers, and Passive Optical Networks (PON) providing high-speed broadband connectivity. However, leveraging PON hardware infrastructure for edge computing presents a promising opportunity to create a high-performance and cost-effective platform for running edge services \cite{tim_edge_cloud, telefonica_whitepaper}. 

% Security risks in the integration
Unlike centralized cloud environments, deploying workloads on PON infrastructures introduces security risks. A primary concern is the physical exposure of hardware in uncontrolled environments, increasing the risk of tampering and unauthorized access \cite{Xiao2019EdgeCS, gai2021attacking}. Security risks are amplified by the complexity of software architectures, which rely on Commercial Off-the-Shelf (COTS) and Open-Source Software (OSS) components for virtualization, software-defined networking, and orchestration. Software reuse provides flexibility and cost-efficiency, but can also introduce software vulnerabilities and expose to compromised software dependencies \cite{ladisa2023sok}. Multi-tenancy introduces more security risks, as different edge applications share the same infrastructure. Maintaining isolation between tenants is critical to prevent escalation of security attacks \cite{varadharajan2016securing}.

% GENIO project (Edge cloud platform enabled by a new intelligent OLT for GPON networks)
The GENIO project \cite{system_management_genio} is a joint R\&D initiative between academic and industry partners, aiming to achieve a secure-by-design edge computing platform integrated with PON networks. Unlike conventional edge models that rely on dedicated servers, GENIO leverages existing PON hardware to host multi-tenant edge services, directly within the telecom infrastructure. This approach can optimize resource utilization and create new revenue opportunities for telecom operators, without the need for additional investments in dedicated servers. One of the main objectives of the GENIO project is to align the platform with security regulations, such as the European Cyber Resilience Act \cite{CRA_eu} and CE marking certification \cite{CE_marking}. This objective shaped the platform by guiding threat mitigations. 

% Limitation of existing security methodologies
Despite the growing emphasis on security-by-design, existing frameworks are not directly applicable for the design of the GENIO platform. Technical standards, e.g., from CISA \cite{cisa} and NIST \cite{nist}, provide high-level security guidelines, but these lack practical, actionable implementation details. In real-world deployments, security must be tailored to the specific hardware and software technology and to the operational constraints of heterogeneous industrial environments, thus requiring customized solutions. Academic research has proposed security-by-design methodologies for specific types of systems, such as cloud applications \cite{casola2020novel}, smart grids \cite{aranha2019enabling}, big data frameworks \cite{awaysheh2021security}, and 5G networks \cite{dutta20205g}. However, no blueprint is readily available for security-by-design in PON-based edge computing. 

% Contribution
This paper presents our findings from the design of the GENIO platform, covering security risks across hardware, OS, middleware, and applications. We evaluated the maturity and limitations of security mitigations, based on OSS solutions, and documented several challenges due to architectural constraints and software heterogeneity. By aligning security research with deployment realities, this work provides key lessons learned, bridging the gap between theoretical security models and practical implementation in an industrial context.
% ---------------------------------------------------------------%

%\bigbreak
\section{The GENIO Project}
The GENIO project aims to develop a platform that integrates edge computing capabilities into telecom central offices, leveraging PON equipment. The architecture spans three layers: the cloud layer, the edge layer, and the far-edge layer, enabling flexible application deployments based on latency and computation requirements.

\begin{figure}[t]
  \centering
  \includegraphics[width=0.6\linewidth]{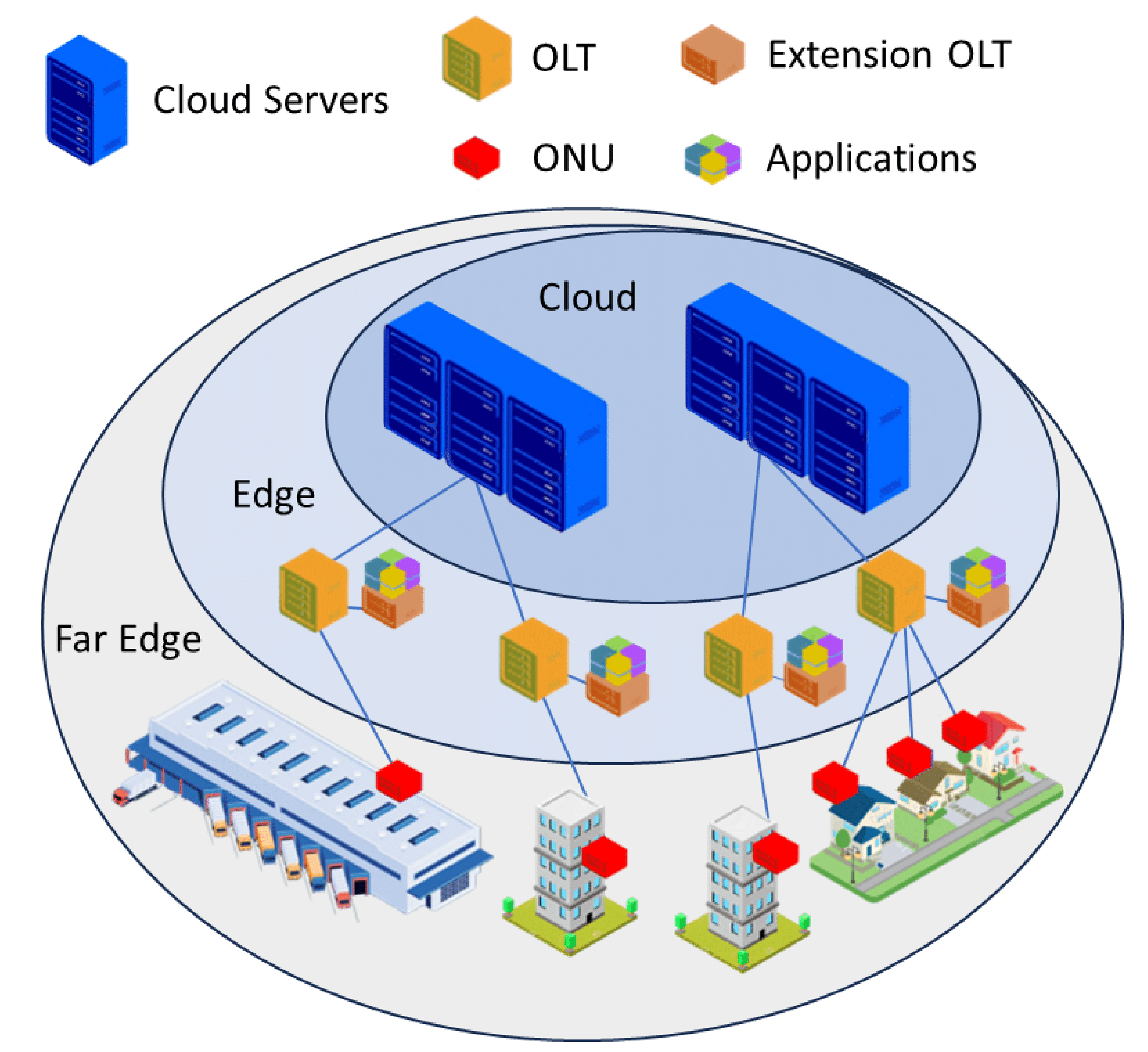}
  \caption{GENIO deployment across cloud, edge and far-edge layers.}
  \label{fig:genio_architecture}
\end{figure}

\noindent
\textbf{Deployment}. As shown in Figure \ref{fig:genio_architecture}, the \textit{far-edge layer} includes Optical Network Units (ONUs), which are deployed in residential and business premises, and which connect users to the fiber network. In GENIO, ONUs are equipped with additional low-end computing resources, enabling them to run applications with ultra-low latency requirements. 
The \textit{edge layer} includes Optical Line Terminals (OLTs), which are devices located in telecom central offices, and which are repurposed in GENIO to serve as edge computing hubs. Originally designed for PON connectivity management, OLTs are enhanced with additional hardware and software to provide computational and storage resources. This layer is optimized for applications with strict latency and bandwidth requirements, balancing performance and resource availability. For applications with less stringent latency requirements, the \textit{cloud layer} offers high computational and storage resources. The cloud layer also behaves as the orchestration center, managing resources across the edge and far-edge layers, and handling complex tasks that exceed the capabilities of local devices. 

\vspace{2mm}
\noindent
\textbf{Use cases}. GENIO is designed to serve both \emph{business users} and \emph{end-users}. 
Business users, such as service operators and enterprises, are providers of edge applications to be shared as container images on a public registry of the GENIO project. Examples include ML workloads, real-time analytics, IoT data processing (e.g., from smart meters, cameras, and sensors), and telecom network functions. Business users can leverage the GENIO platform through an Infrastructure-as-a-Service (IaaS) model, by leasing computing, storage, and networking resources on the edge to run their applications. End-users, which include individual customers and businesses, interact with the platform via a Software-as-a-Service (SaaS) model, by consuming edge applications provided by business users.

\vspace{2mm}
\noindent
\textbf{Software Architecture}. The GENIO platform a distributed and multi-layered architecture to manage network operations, resource allocation, and application deployment. In particular, OLTs are based on x86 COTS hardware, and they integrate several OSS technologies for software-defined networking (SDN), including ONOS \cite{onos}, VOLTHA \cite{voltha_doc} and Open Networking Linux (ONL) \cite{onl}, to dynamically and efficiently manage PON network resources. OLTs are extended to support secure multi-tenancy for running edge applications, based on OSS virtualization technologies. The physical resources of the OLT are managed using a cluster of virtual machines, managed using the Linux/KVM hypervisor. Edge applications can run in either hard isolation (in dedicated virtual machines) or soft isolation (in containers and network namespaces within the virtual machines), to accomodate different security and performance requirements. Additionally, GENIO uses Kubernetes and Proxmox \cite{proxmox} for orchestrating the virtual machines and containerized applications, allowing scalable, resilient, and efficient workload management based on current network and computational conditions. 

%...................................................................................:%

\section{Threat Modeling for GENIO}
\label{sec:threat_model}
\begin{figure}[!t]
  \centering
  \includegraphics[width=0.5\textwidth]{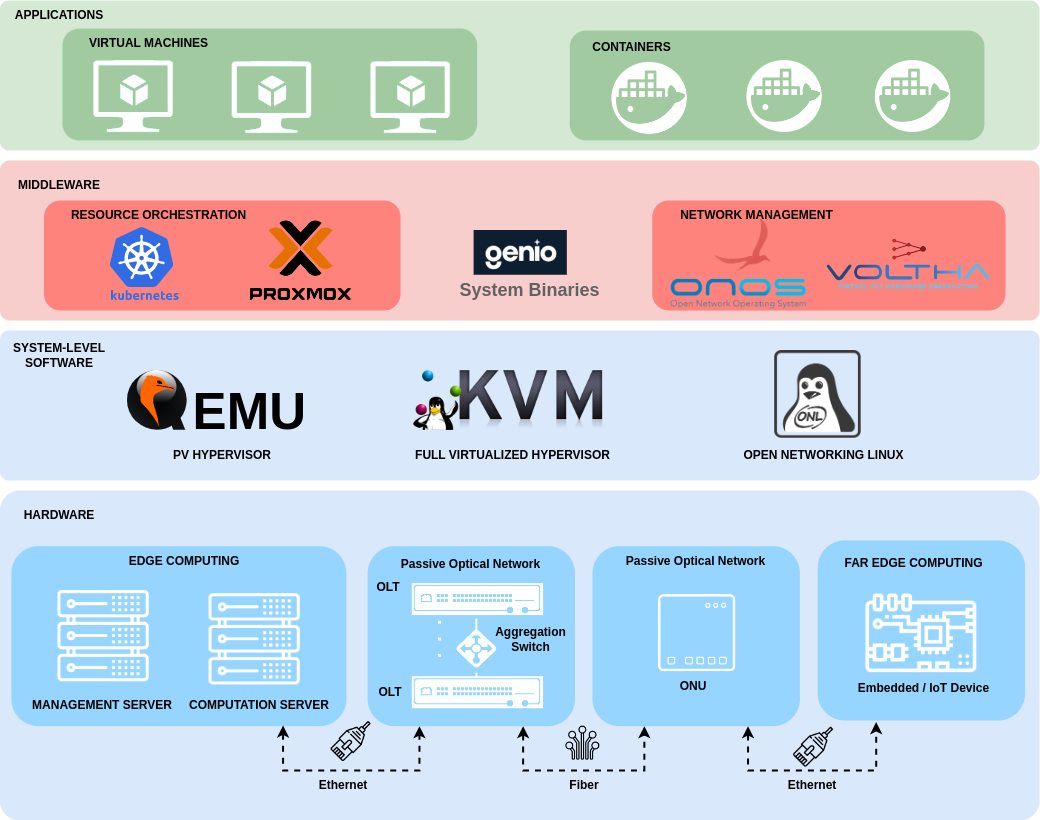}
  \caption{GENIO architecture.}
  \label{fig:threat_model}
\end{figure}

\begin{figure}[t]
  \centering
  \includegraphics[scale=0.35]{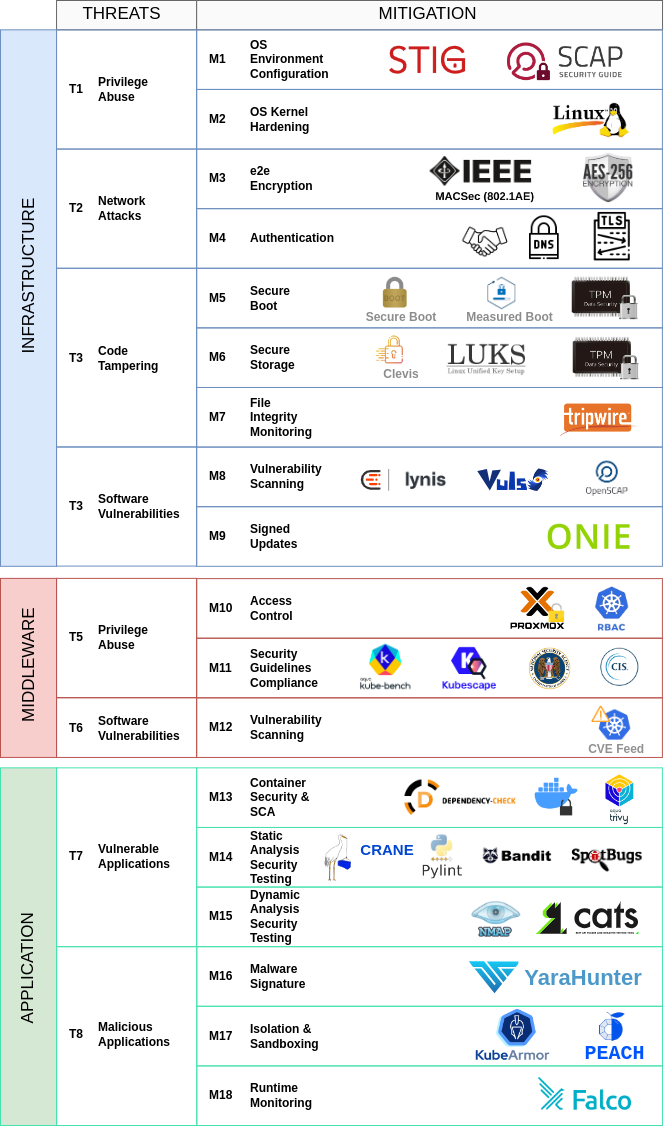}
  \caption{OSS security solutions and standards in GENIO.}
  \label{fig:security_mitigations}
\end{figure}

Securing the GENIO platform required developing a comprehensive threat model to identify risks across the cloud, edge, and far-edge layers. Using the STRIDE methodology \cite{stride}, we systematically identified threats, from physical tampering of ONUs and OLTs, to software vulnerabilities and misconfigurations of orchestration services. This process led to the categorization of risks into \emph{Infrastructure-level}, \emph{Middleware-level}, and \emph{Application-level} threats. Figures~\ref{fig:threat_model} and \ref{fig:security_mitigations} provide an overview of the GENIO architecture and a summary of threats and mitigations, respectively.

\subsection{Infrastructure-level Threats}
The infrastructure level encompasses hardware components and low-level software, which are the foundation of the GENIO architecture.

\vspace{2mm}
\noindent
\threatlabel{\textcolor{red!30!black}{T1} Network Attacks}
The distributed nature of the GENIO architecture creates multiple points of vulnerability for secure data transmission, spanning OLTs, ONUs, inter-OLT links, and cloud interactions. Adversaries can eavesdrop, modify traffic, or impersonate network components at various stages, with \emph{interception and replay attacks} posing a direct threat to data integrity and authenticity. Physical exploits like \emph{downstream hijacking} and \emph{ONU impersonation} target the PON architecture at the hardware and firmware level. Infrastructure-level tampering often involves \emph{physically tapping fiber connections} \cite{iqbal2011optical} or modifying device firmware to \emph{siphon traffic} \cite{siphon-traffic}.

\vspace{2mm}
\noindent
\threatlabel{\textcolor{red!30!black}{T2} Code Tampering}
Attackers can target low-level system components to introduce persistent threats, embedding malware or backdoors into the platform. \emph{Reverse engineering}, \emph{binary untrusted patching/updating}, and \emph{firmware manipulation} are common attack techniques that allow adversaries to manipulate hypervisors, kernels, and system binaries. A successful compromise at this level can provide long-term control over the entire host machine. 

\vspace{2mm}
\noindent
\threatlabel{\textcolor{red!30!black}{T3} Privilege Abuse}
Misconfigurations in low-level software, such as unrestricted OS accounts, services, and files, can expose the system to \emph{privilege escalation}. Intruders can exploit these flaws to expand their control over the platform, by hijacking administrative functions and achieving persistency. This can facilitate service disruptions and data thefts, posing a severe risk to the security of GENIO operations.

\vspace{2mm}
\noindent
\threatlabel{\textcolor{red!30!black}{T4} Software Vulnerabilities}
Unpatched or unknown vulnerabilities in low-level software can be exploited by attackers to gain full access to the host machine, and to break isolation mechanisms. Unfortunately, handling these vulnerabilities can be quite difficult, since OLTs and ONUs are managed and updated remotely. Moreover, the GENIO platform relies on a custom Linux kernel configuration to support SDN software, requiring continuous vulnerability assessment to secure its custom stack. Any failure to address these vulnerabilities can expose the infrastructure to \emph{kernel exploits} and \emph{container escaping}. %, and \emph{hypervisor takeovers}.

%...................................................................................:%
\subsection{Middleware-level Threats}
\label{subsec:middleware}
The middleware level in GENIO includes software-defined networking (VOLTHA, ONOS), virtualization and container management (Proxmox, Kubernetes). These components provide powerful interfaces to programmatically manage resources, but also introduce more security challenges.

\vspace{2mm}
\noindent
\threatlabel{\textcolor{red!30!black}{T5} Privilege Abuse}
\emph{Misconfigurations} can also apply to middleware, such as overprivileged roles and unrestricted API access. Weak Role-Based Access Control (RBAC) policies can grant excessive permissions, enabling privilege escalation and lateral movement. This risk is often exacerbated by \emph{insecure defaults} in open-source software \cite{insecure_defaults} \cite{insecure_defaults_cwe}, which prioritizes usability over security by not enabling strict execution policies and strong authentication mechanisms. Without proper hardening, attackers can exploit misconfigurations to manipulate workloads, gain unauthorized access, and disrupt network operations. 

\vspace{2mm}
\noindent
\threatlabel{\textcolor{red!30!black}{T6} Software Vulnerabilities}
Software vulnerabilities can also arise from flaws in orchestration and network management software, which represent a significant share of the codebase. These weaknesses, such as \emph{bugs in workflows and API implementations} \cite{ju2013openstack,cotroneo2019openstack} and \emph{vulnerable third-party dependencies} \cite{dependency-third-party-cwe}, can be exploited to compromise middleware security. These these flaws expose middleware resources to unintended access, allowing adversaries to intercept sensitive data. 

%...................................................................................:%

\subsection{Application-level Threats}
\label{subsec:application}
The GENIO platform supports the deployment and execution of applications across its far-edge, edge, and cloud layers. Applications can expose other ones, and the GENIO platform itself, to security attacks.

\vspace{2mm}
\noindent
\threatlabel{\textcolor{red!30!black}{T7} Vulnerable Applications}
Since applications are delivered by third-party business users, they can bring additional \emph{application vulnerabilities}. Attackers can exploit such vulnerabilities to gain foothold on a tenant, and pursue malicious actions against users, other tenants, and the underlying platform. Application vulnerabilities arise from the lack of secure software development practices, such as static/dynamic analysis and reuse of insecure components. These issues can expose users to data breaches (e.g., through SQL injection) and injection attacks (e.g., Cross-Site Scripting). Moreover, attackers can gain access through command injection, deserialization, and memory corruption vulnerabilities, which can lead to remote code execution.

\vspace{2mm}
\noindent
\threatlabel{\textcolor{red!30!black}{T8} Malicious Applications}
Malicious behaviors can arise both from exploited vulnerabilities (as previously discussed), and from \emph{deliberately malicious applications}. For example, business users can reuse malicious container images from external repositories, which can contain hidden malware or backdoors. These untrusted applications can bypass scrutiny through obfuscation. These applications can execute malicious code to invoke privileged system calls and misusing capabilities (e.g., \texttt{CAP\_SYS\_ADMIN} in Linux containers), in order to escape container restrictions and disrupt the host and neighboring services. 
Moreover, malicious applications can attack the platform through \emph{resource abuse}, by monopolizing CPU, memory, network, and storage resources, thereby degrading performance and causing service outages for other tenants. 

% ---------------------------------------------------------------%
%\section{Mitigations}
%This section discusses security mitigations using OSS across the infrastructure, middleware, and application layers. 
%We analyze their maturity when used in an industrial setting like GENIO, highlighting challenges and lessons learned from real-world deployment. 
%Figure~\ref{fig:security_mitigations} summarizes the mitigations, with related OSS solutions. 

\section{Infrastructure-level Mitigations}

\subsection{Mitigating Privilege Abuse}
\noindent
\mitigation{\textcolor{green!30!black}{M1} OS environment configurations} 
GENIO ensures a secure ONL Linux configuration, by stripping non-essential components (e.g., unused packages, services, kernel modules) to minimize the attack surface. Security policies are automated using OpenSCAP \cite{openscap}, which enforces SCAP benchmarks to secure SSH configurations, enable NTP synchronization, disable untrusted APT repositories, and protect kernel files. GENIO also aligns to Security Technical Implementation Guides (STIGs) \cite{stig}, a set of best practices originally developed for Ubuntu and other mainstream Linux distributions, to enforce encryption policies, restrict system access, and secure boot configurations.

\vspace{2mm}
\noindent
\mitigation{\textcolor{green!30!black}{M2} OS kernel hardening} At the kernel level, memory protections (e.g., \texttt{CONFIG\_STACKPROTECTOR}) block buffer overflow attacks, while Linux Security Modules (AppArmor/SELinux) \cite{linux-security-modules} restrict privileged system calls. High-risk functionalities like \texttt{KEXEC} (runtime kernel replacement) and \texttt{KPROBES} (debugging hooks) are disabled. The kernel-hardening-checker tool \cite{kernel-hardening-checker} validates configurations (kconfig, cmdline, sysctl) against hardened baselines, and speculative execution mitigations (Intel/AMD microcode \cite{spectre}) address side-channel vulnerabilities like Spectre \cite{Kocher2018spectre}.

\Lesson{
The platform’s reliance on Open Networking Linux (ONL) introduced complexities, as ONL lacks formal security guidelines compared to mainstream distributions. The application of STIGs and SCAP benchmarks was required to align with ONL’s architecture, demanding iterative adjustments and reviews to balance security, performance, and compatibility.
}

\subsection{Securing Communication}
\noindent
\mitigation{\textcolor{green!30!black}{M3} End-to-End Encryption} GENIO safeguards traffic across both Ethernet and PON segments through end-to-end encryption to prevent interception or tampering. At layer 2, the MACsec protocol standardized by \textit{IEEE 802.1AE} \cite{ieee-802.1ae} encrypts raw Ethernet frames using AES-GCM, providing authentication, confidentiality, and integrity for data on point-to-point Ethernet. In parallel, GENIO follows optical-specific guidelines such as \textit{ITU-T G.987.3} \cite{itu-t-g.987.3} for GPON, which recommend AES-based payload encryption to defend against fiber taps and low-level tampering in PONs.

\vspace{2mm}
\noindent
\mitigation{\textcolor{green!30!black}{M4} Authentication of Nodes} GENIO enforces mutual authentication between ONUs and OLTs to verify hardware legitimacy before service provisioning. Certificate-based methods (via PKI) validate device identities, preventing rogue devices from impersonating legitimate infrastructure. Secure key exchange protocols (e.g., TLS 1.3) and secure DNS \cite{dns_security} prevent man-in-the-middle attacks during onboarding and registration, ensuring only trusted devices access network resources.

\Lesson{
Encryption imposes additional engineering efforts and computational resources to enhance the security of the PON network. 
Implementing secure authentication among heterogeneous hardware (ONUs, OLTs, and cloud systems) demands careful management of certificates. GENIO’s alignment with evolving ETSI standards \cite{etsi-ts-103-963} reflects an ongoing effort to maintain interoperability and compliance with industry guidelines.
}
 
\subsection{Ensuring Code Integrity}
\noindent
\mitigation{\textcolor{green!30!black}{M5} Secure Boot} GENIO uses \textit{Secure Boot} and a \textit{Trusted Platform Module} (TPM) to cryptographically verify OS and firmware components before execution. At the earliest stage, the \textit{Shim} \cite{shim} bootloader, signed by a recognized certificate authority (e.g., Microsoft), initializes a secure environment before loading the GRUB bootloader. By relying on Shim, the GENIO platform can add custom keys to validate later boot layers, including distribution-specific kernels. Additionally, \textit{Measured Boot} records hashes of critical binaries in TPM Platform Configuration Registers (PCRs) at boot, enabling integrity checks against expected values. Together, these measures help ensure the platform boots from a known-good state and reveal any subsequent compromise.

\vspace{2mm}
\noindent
\mitigation{\textcolor{green!30!black}{M6} Secure Storage} Beyond TPM-based firmware and OS verification, GENIO protects data at rest using \textit{Linux Unified Key Setup} (LUKS) \cite{luks} to encrypt entire partitions with a passphrase. After encryption, the decryption key can be bound to specific PCR values in the TPM. If the measured environment (e.g., the kernel) matches the expected hash chain, the TPM releases the decryption secret; otherwise, access is denied. To automate this, GENIO plans to integrate \textit{Clevis} \cite{clevis}, which seamlessly unwraps the LUKS key at boot when TPM-managed PCRs confirm system integrity. This enables booting without manual passphrase entry, reducing operational overhead in PON settings.

\vspace{2mm}
\noindent
\mitigation{\textcolor{green!30!black}{M7} File Integrity Monitoring} Even with secure boot, adversaries may attempt to alter system files post-boot. GENIO deploys \textit{Tripwire} \cite{tripwire} for runtime file integrity monitoring (FIM), creating cryptographic baselines of critical system files and alerting administrators to unauthorized changes. Tripwire’s configurations and databases are encrypted and signed, with keys protected by the TPM to prevent tampering with the monitoring process.

\Lesson{
Deploying integrity protections in industrial environments faces obstacles. GENIO relies on older OS distributions (ONL Linux, based on Debian 10) to run SDN software and drivers, which lack native support for recent software packages. As a result, manually installing newer dependencies is required, introducing potential conflicts. Libraries required by Clevis for TPM access and automated disk decryption are unavailable, forcing manual passphrase entry at boot, which is impractical for in-field deployments of OLT nodes. Moreover, file monitoring should distinguish between critical resources that should not be mutable (e.g., system binaries and configurations) from mutable ones, to avoid misleading alerts.
}

\subsection{Mitigating Software Vulnerabilities}
\noindent
\mitigation{\textcolor{green!30!black}{M8} Automated Scanning} GENIO conducts periodic vulnerability scanning with tools such as OpenSCAP \cite{openscap}, Lynis \cite{lynis} and Vuls \cite{vuls} to detect known CVEs across the low-level software, including the Linux kernel, system binaries, and user-space packages. These reports are prioritized based on severity and exploitability, ensuring that critical patches are applied as soon as feasible. 

\vspace{2mm}
\noindent
\mitigation{\textcolor{green!30!black}{M9} Signed Updates} Ensuring the authenticity and integrity of software updates is critical to thwarting supply-chain attacks within GENIO. The platform thus employs different methods tailored to each update scenario. In Debian-based environments, \textit{user-space packages} are distributed via APT, which signs metadata and packages with GPG keys for each repository, and rejects any unverified artifacts. In addition, GENIO employs ONIE \cite{onie} for securely delivering ONL kernel updates. Following NIST SP 800-193 \cite{NIST-sp-800-193} guidelines, ONIE images are signed with X.509 certificates, accompanied by a detached signature file that is validated against a locally trusted public key, backed by a TPM. ONIE reboots the system into a minimal environment to apply the update, and fully run this environment by using Secure Boot, reducing potential inference from a compromised OS. Beyond kernel and user-space package updates, GENIO must also distribute additional binaries, such as specialized daemons and custom tools. These are also signed with GENIO’s own certificates, which are likewise validated on each target node before installation.
%\hl{A framework for securing software update systems https://theupdateframework.io/}

\Lesson{
The maturity of automated scanning solutions facilitated smooth integration into GENIO’s custom stack, even if occasional manual tuning is required to handle non-standard paths and configurations in ONL. APT GPG signatures for Debian-based images represent a reliable and straightforward solution to adopt. 
} 

% ---------------------------------------------------------------%

\section{Middleware-level Mitigations}

\subsection{Mitigating Privilege Abuse}
\noindent
\mitigation{\textcolor{green!30!black}{M10} Access Control}
GENIO applies the principle of least privilege across its middleware stack, ensuring each role and service holds only the permissions necessary for legitimate operations. For virtualization and container management, native access control frameworks \cite{rbac-kubernetes,proxmox_access_control} can mitigate abuses of resources exposed through their APIs. In network management software, including ONOS and VOLTHA, built-in authentication and authorization mechanisms \cite{onos-security} are configured to prevent API misuse. Exposed APIs for OLT and ONT management are strictly restricted to administrative service accounts secured by TLS certificates. On the network-side, GENIO enforces a clearly defined set of capabilities required in production, such as device registration, logical network configuration, and diagnostic logging—while blocking operations that introduce unnecessary privilege risks, such as direct shell access, low-level debugging endpoints, or raw log retrieval.

\vspace{2mm}
\noindent
\mitigation{\textcolor{green!30!black}{M11} Security Guideline Compliance}
GENIO adheres to industry-recognized security standards and continuously audits configurations to maintain compliance. It implements the NSA Kubernetes Hardening Guidance \cite{nsa-kubernetes-hardening}, CIS Benchmarks \cite{cis-benchmarks}, and a suite of community tools, including docker-bench \cite{docker-bench}, kube-bench \cite{kube-bench}, kubesec \cite{kubesec}, kube-hunter \cite{kube-hunter}, and kubescape \cite{kubescape} to detect misconfigurations in Kubernetes clusters. Additionally, it follows vendor-specific guidelines for SDN-controllers provided by ONOS \cite{onos_security_1} to address insecure defaults, enforce strong authentication, and detect configuration drift.

\Lesson{
Hardening network management software is straightforward, as required capabilities are well-defined, and unnecessary functions can be blocked without disruption. In contrast, the configuration of RBAC policies for the orchestration platforms is challenging, since they are feature-rich and resource access should be carefully adapted for the workflows of the GENIO platform. Moreover, designers must integrate multiple security guidelines and checker tools, since individual solutions only address a subset of the risks.
}

\subsection{Mitigating Software Vulnerabilities}
\noindent
\mitigation{\textcolor{green!30!black}{M12} Automated Scanning and Patching} 
GENIO integrates multiple sources to track vulnerabilities in middleware components. For Kubernetes, it leverages its official CVE database \cite{cve-kubernetes}, which provides real-time updates on disclosed vulnerabilities, affected versions, exploitability, impact, and patches. The Kubernetes database offers a structured, programmatically accessible CVE feed for automated monitoring. Other middleware components vary in vulnerability tracking maturity. The Docker runtime publishes security updates \cite{docker-forum} as blog-format announcements, making structured extraction difficult. ONOS maintains a structured web interface but is no longer actively updated. Proxmox notifies users only via its web UI. For middleware lacking structured, up-to-date, or programmatically accessible feeds, GENIO relies on the National Vulnerability Database (NVD) APIs to track vulnerabilities. To enhance precision in Kubernetes vulnerability tracking, GENIO integrates the Kubernetes Bill of Materials (KBOM) \cite{kbom}, which catalogs control plane services, node components, and add-ons with their exact versions and images, mapping known vulnerabilities in installed components.

\Lesson{
Middleware vulnerability management remains reactive and resource-intensive, since tracking vulnerabilities involves fragmented sources. 
The NVD API, despite its completeness, still requires manual reviews. 
The owner of the platform must cross-reference security advisories with deployed versions, assess exposure, and schedule patches—delays that extend the attack window in production environments. 
}
 
% ---------------------------------------------------------------%

\section{Application-level Mitigations}
\subsection{Mitigating Software Vulnerabilities}
\noindent
\mitigation{\textcolor{green!30!black}{M13} Container Security and SCA}
A key aspect of securing applications in GENIO is hardening containerized workloads. The platform uses Docker Bench for Security \cite{docker-bench} to detect and fix misconfigurations that may introduce vulnerabilities. By enforcing best practices, such as least-privilege execution, restricted volume mounting, and secure networking, GENIO reduces the application attack surface. To address risks from third-party dependencies, GENIO integrates Software Composition Analysis (SCA) with tools like Trivy \cite{trivy} and OWASP Dependency Check \cite{owasp-checker}, scanning container images, identifying imported packages, and matching versions against CVE databases. These tools provide visibility into vulnerabilities introduced by pre-built components in the application stack.

\vspace{2mm}
\noindent
\mitigation{\textcolor{green!30!black}{M13} Static Application Security Testing}
GENIO applies SAST to detect vulnerabilities in the source code of the application itself. The container filesystem is extracted using Crane \cite{crane}, allowing further scans to detect quality issues. Java source files are analyzed with SpotBugs \cite{spotbugs}, identifying issues such as null pointer dereferences, improper resource management, and inefficient exception handling. Pylint \cite{pylint} serves as the Python counterpart. In addition, GENIO integrates Semgrep \cite{semgrep} and Bandit \cite{bandit} to detect security vulnerabilities, such as hardcoded credentials, improper input validation, and weak cryptographic functions.

\vspace{2mm}
\noindent
\mitigation{\textcolor{green!30!black}{M15} Dynamic Application Security Testing}
GENIO integrates DAST to uncover runtime vulnerabilities. Specifically, GENIO employs CATS \cite{cats}, a REST API fuzzer tool, to evaluate OpenAPI-defined endpoints. CATS conducts fuzz testing by injecting malformed, unexpected, and malicious inputs to identify vulnerabilities such as insecure input handling, improper authentication enforcement, and API misconfigurations. Beyond application-level testing, GENIO enforces network security checks when the application is deployed. Nmap \cite{nmap} verifies TLS enforcement to ensure secure communication and analyzes port configurations, identifying unnecessary open ports that could expose to external threats.

\Lesson{
While SCA and SAST tools are mature and widely available, integrating them into GENIO poses challenges. SCA often flags unused or misidentified dependencies, generating noise and false positives. It also analyzes entire dependencies without linking vulnerabilities to specific functions used by the application, resulting to bloated reports and complicating path prioritization. Lastly, fuzzing containerized applications is feasible only for those exposing standard interfaces, such as REST APIs.
}

\subsection{Identifying Malicious Applications}
\noindent
\mitigation{\textcolor{green!30!black}{M16} Malware Signature}
GENIO defends against malicious applications using malware signature detection to proactively identify known malicious components before they are deployed or executed. To this end, GENIO utilizes Deepfence YaraHunter \cite{yarahunter} to scan container images at rest for indicators of compromise. This tool leverages YARA rules to detect embedded malicious binaries, scripts, or configuration files.

\vspace{2mm}
\noindent
\mitigation{\textcolor{green!30!black}{M17} Isolation \& Sandboxing} 
To limit the impact of malicious applications, GENIO enforces strict runtime boundaries through isolation and sandboxing. It integrates KubeArmor \cite{kubearmor} to restrict container, pod, and VM behavior at the system level using Linux Security Modules (LSMs), blocking unauthorized processes, file access, and suspicious network activity. To strengthen multi-tenancy isolation, GENIO follows the best practices from the PEACH framework \cite{peach}, which models isolation risks based on interface complexity, tenant separation, and enforcement strength across key dimensions such as privilege, encryption, and authentication. 
%GENIO uses this analysis to apply compensating controls that reduce the impact of potential cross-tenant vulnerabilities.

\vspace{2mm}
\noindent
\mitigation{\textcolor{green!30!black}{M18} Runtime Monitoring}
To enhance visibility into the behavior of running applications, GENIO integrates Falco \cite{falco}. Falco monitors system calls in real-time using eBPF and evaluates them against a rich, customizable rule set to detect suspicious behaviors such as unexpected shell execution, unauthorized file access, or unusual network connections. Unlike signature-based scanners or sandboxers, Falco provides deep runtime observability without blocking execution, enabling early detection of post-exploitation activities.

\Lesson{
Our experience with these tools confirms that these techniques are relatively mature and effective in detecting and isolating malicious applications. However, challenges remain in tuning policies and rules to minimize false positives without weakening security. Maintaining performance overheads within acceptable bounds is also a key consideration. 
}

\section{Conclusion}
Securing an edge computing platform demands to carefully address threats across all layers of the system. The GENIO project identifies and addresses key risks spanning infrastructure, middleware, and applications. Mitigation strategies were deployed using open-source tools and best practices, revealing both the strengths and limits of current security solutions.

\section*{Acknowledgments}
This project has been partially supported by the GENIO project (CUP B69J23005770005) funded by MIMIT, "Accordi per l'Innovazione" program.

\balance
\bibliographystyle{IEEEtran}
%\bibliography{bibliography}

\end{document}